\documentclass[reprint, prl, twocolumn, superscriptaddress, floatfix, showpacs, altaffilletter]{revtex4-1}

\usepackage{graphicx}
\usepackage{amsmath}
\usepackage{amsfonts}
\usepackage{mathrsfs}
\usepackage[percent]{overpic}
\usepackage{subfigure}
\input{babarsym}

\def\Nbar    {\kern 0.18em\overline{\kern -0.18em N}{}\xspace}
\def\Nb      {\ensuremath{\Nbar}\xspace}
\def\NN      {\ensuremath{N\Nb}\xspace}

\allowdisplaybreaks[1]

\newcommand{\twoSresult}{\ensuremath{(2.64 \pm 0.11\stat {}^{+0.26}_{-0.21}\syst)\! \times \! 10^{-5}}}
\newcommand{\threeSresult}{\ensuremath{(2.33 \pm 0.15\stat {}^{+0.31}_{-0.28}\syst)\! \times \! 10^{-5}}}
\newcommand{\oneSresult}{\ensuremath{(2.81 \pm 0.49 \stat {}^{+0.20}_{-0.24}\syst)\! \times \! 10^{-5}}}
\newcommand{\contresult}{\ensuremath{(9.63 \pm 0.41\stat {}^{+1.17}_{-1.01}\syst) \fb }}

\newcommand{\twoSresultSm}{\ensuremath{(2.64 \pm 0.11 {}^{+0.26}_{-0.21})\! \times \! 10^{-5}}}
\newcommand{\threeSresultSm}{\ensuremath{(2.33 \pm 0.15 {}^{+0.31}_{-0.28})\! \times \! 10^{-5}}}
\newcommand{\oneSresultSm}{\ensuremath{(2.81 \pm 0.49 {}^{+0.20}_{-0.24})\! \times \! 10^{-5}}}
\newcommand{\contresultSm}{\ensuremath{(9.63 \pm 0.41 {}^{+1.17}_{-1.01}) \fb }}

\newcommand{\twoSresultLong}{\ensuremath{\mathcal{B}(\Y2S \to \bar{d}X) = \twoSresult}}
\newcommand{\threeSresultLong}{\ensuremath{\mathcal{B}(\Y3S \to \bar{d}X) = \threeSresult}}
\newcommand{\oneSresultLong}{\ensuremath{\mathcal{B}(\Y1S \to \bar{d}X) = \oneSresult}}
\newcommand{\contresultLong}{\ensuremath{\sigma (\epem \to \bar{d}X)=\contresult}}

\newcommand{\continuum}{\ensuremath{\epem \to \qqbar}\xspace}
\newcommand{\YnS}{\ensuremath{\Upsilon(nS)}\xspace}

\begin{document}
\title{
\begin{flushleft}
       \mbox{\textmd{ \normalsize \babar-PUB-13/020 }}
       \\ \mbox{\textmd{ \normalsize SLAC-PUB-15925 }}
       \end{flushleft}
{
\large \boldmath
Antideuteron production  in \Y{n}S  decays and in \continuum
at $\sqrt{s} \approx 10.58 \gev$}
}

%
\author{J.~P.~Lees}
\author{V.~Poireau}
\author{V.~Tisserand}
\affiliation{Laboratoire d'Annecy-le-Vieux de Physique des Particules (LAPP), Universit\'e de Savoie, CNRS/IN2P3,  F-74941 Annecy-Le-Vieux, France}
\author{E.~Grauges}
\affiliation{Universitat de Barcelona, Facultat de Fisica, Departament ECM, E-08028 Barcelona, Spain }
\author{A.~Palano$^{ab}$ }
\affiliation{INFN Sezione di Bari$^{a}$; Dipartimento di Fisica, Universit\`a di Bari$^{b}$, I-70126 Bari, Italy }
\author{G.~Eigen}
\author{B.~Stugu}
\affiliation{University of Bergen, Institute of Physics, N-5007 Bergen, Norway }
\author{D.~N.~Brown}
\author{L.~T.~Kerth}
\author{Yu.~G.~Kolomensky}
\author{M.~J.~Lee}
\author{G.~Lynch}
\affiliation{Lawrence Berkeley National Laboratory and University of California, Berkeley, California 94720, USA }
% ...
\collaboration{The \babar\ Collaboration}
\noaffiliation

\begin{abstract}
We present measurements of the inclusive production of antideuterons in \epem annihilation into hadrons 
at $\approx 10.58$\gev center-of-mass energy and in $\Upsilon(1S,2S,3S)$ decays. 
The results are obtained using data collected by the \babar\ detector at the \pep2\ electron-positron collider. 
Assuming a fireball spectral shape for the emitted antideuteron momentum, 
we find \oneSresultLong, \twoSresultLong, \threeSresultLong, and \contresultLong.
\end{abstract}

\maketitle

The production of nuclei and anti-nuclei in hadronic collisions and in hadronization processes has 
recently attracted considerable theoretical and experimental interest~\cite{Cui:2010ud,Dal:2012my,Vittino:2013qna},
since cosmic anti-nuclei may provide a sensitive probe of dark matter annihilation.
Dark matter particles might annihilate into two colored partons -- quarks ($q$) and gluons ($g$) -- 
which could hadronize into mesons and baryons, potentially forming bound states such as light (anti)nuclei.
The latter process, requiring at least six $q$ or $\qbar$ in close proximity, is
poorly understood both theoretically and experimentally, and precise
measurements of both total rates and momentum spectra are needed.
With no initial-state hadrons, $\epem$ annihilations provide a clean
probe of this process not only for $q$ and $\qbar$ but also for $g$ via
decays of $\Upsilon$ and other vector resonances.

Experimental measurements focus on antideuteron (\dbar)
production as such studies are not limited, as in the deuteron (\d)
production case, by the high rate of nuclei production via interactions with the detector materials. 
The ARGUS~\cite{Albrecht:1989ag} and CLEO~\cite{Asner:2006pw} experiments observed \dbar production at the level of
$3 \times 10^{-5}$ per \OneS and \TwoS decays, and set limits on production in \FourS
decays and \continuum at 10.6 \gev.  
The ALEPH~\cite{Schael:2006fd} experiment observed a 3$\sigma$ evidence for \dbar production in \continuum at 91.2 \gev.
In these measurements, the accessible kinematic range, 0.4 -- 1.7 \gevc, 
was representing less than 20\% of the phase space where \d and \dbar could be identified.

In this Letter, we present studies of \dbar production in \epem
annihilation data taken on and just below the \TwoS, \ThreeS and \FourS resonances. To avoid any ambiguity, we
refer to \d and \dbar separately everywhere in this note, so that charge conjugation is not implied anywhere. 
We also study the \dbar production from \OneS using the $\TwoS \to \OneS \pip \pim$ decay chain.  
The boost of the center of mass (CM) allows a wide momentum range to be accessed,
0.3 -- 3 \gevc, which corresponds to 0.5 -- 1.5 \gevc in the laboratory frame.
We confirm the \OneS rates, improve the coverage and precision for \TwoS decays, measure \dbar
production in \ThreeS decays, and also, for the first time, in \continuum near 10.6 \gev.

The results presented here are obtained from the complete \babar\ $\Upsilon(2S,3S,4S)$ (\YnS)  datasets (Onpeak), 
including data collected at a CM energy 40\mev below the peak of each resonance (Offpeak). 
The luminosity~\cite{Lees:2013rw} collected for each dataset and
the corresponding number of \YnS decays are reported in Table~\ref{tab:lumi}.
\begin{table}[bht]
\vspace{-0.3cm}
\caption{Collected luminosity and number of decays for the samples used in the analysis. 
Integrated luminosity is reported both for the Onpeak and Offpeak samples.}
\begin{center}
\begin{ruledtabular}
\begin{tabular}{l|ccc}
{\bf Resonance} & {\bf Onpeak} & {\bf \# of $\Upsilon$ Decays} & {\bf Offpeak} \\
\hline
\FourS & 429\invfb & $463 \times 10^6$ & 44.8\invfb\\
\ThreeS & 28.5\invfb & $116 \times 10^6$ & 2.63\invfb \\
\TwoS & 14.4\invfb & $98.3 \times 10^6$ & 1.50\invfb \\
\end{tabular}
\end{ruledtabular}
\vspace{-0.4cm}
\label{tab:lumi}
\end{center}
\end{table}
We also use Monte Carlo (MC) simulated data samples 
generated using {\sc JetSet}~\cite{JETSET} for $\epem \to \qqbar \, (q=u,d,s)$ events 
and {\sc EvtGen}~\cite{EVTGEN} for \YnS decays. The interaction of simulated particles 
with the \babar\ detector is modeled using {\sc Geant4}~\cite{GEANT}.
Neither \d nor \dbar production is implemented in {\sc JetSet}, and \dbar cannot be simulated
in the version of {\sc Geant4} that is used. 
Therefore we use the {\sc EvtGen} phase-space generator and {\sc Geant4} to simulate 
$\Upsilon(2S,3S) \to d \bar{N}\bar{N'} (5h)$ decays 
(where $h$ indicates a $K^{\pm}$, $\pi^{\pm}$, or $\pi^0$, and $N,N'=p,n$) 
and $\epem \to \qqbar 
\to d \bar{N}\bar{N'} (5h)$ events 
for studying reconstruction efficiencies. 
The additional five hadrons are a representative average of additional particles in the decay and 
restrict the phase-space, so we have \d's with CM momentum lower than 3 \gevc.
These samples will be referred to as ``signal MC'' throughout this Letter, though we note that
they are not expected to reproduce the global features of real signal events nor the distribution
in momentum or polar angle of $d$ or $\bar{d}$ in data. 
To account for differences between data and MC samples, 
corrections are applied and systematic uncertainties are assigned, as discussed below.

The \babar\ detector, trigger, and the coordinate system used throughout, 
are described in detail in Refs.~\cite{Aubert:2001tu, TheBABAR:2013jta}.
The most relevant part of the detector for this analysis is the tracking system, 
composed of a 5-layer inner silicon strip tracker, the Silicon Vertex Tracker (SVT), and the 40-layer
small-cell Drift Chamber (DCH) inside a 1.5 Tesla axial magnetic field. 
The SVT provides information on track parameters near the interaction point (IP),
while the DCH has a 98\% efficiency for detecting charged particles with $p_T > 500\mevc$. 
The $p_T$ resolution is $\sigma_{p_T}/p_{T}=(0.13 \,(\mathrm{Ge\kern -0.1em V\!/}c)^{-1} \cdot p_T + 0.45)\%$. 
The ionisation energy loss ($dE/dx$) is
measured by the two systems, with a resolution of approximately 14\% and
7\%, for the SVT and DCH, respectively.
Additional particle identification information is provided 
by a Detector of Internally-Reflected Cherenkov light 
(DIRC), which, as described later, is employed in this analysis to provide a veto.

Hadronic events are selected by a filter which requires greater than
two reconstructed tracks
and a ratio of second to zeroth Fox-Wolfram
moments~\cite{Fox:1978vu} less than 0.98. 
The reconstructed momentum of the candidate tracks is corrected 
by $0.019 (\gevc)^3 / p^2$
to account for the underestimation of energy loss due to the pion mass assumption employed in the track fit. 
Candidates are retained only if they are within the full polar angle
acceptance of the DCH ($-0.80 \leq \cos{\theta_{\rm LAB}} \leq 0.92$) and
within $0.5 \leq p_{\rm LAB} \leq 1.5 \gevc$, 
where the most probable $dE/dx$ of \d and \dbar is well separated from that of other particle species. 
Here and throughout this Letter ``LAB" denotes observables in the laboratory frame.
To reject candidates with poorly-measured $dE/dx$,
we require that the ionization along the track
trajectory be sampled at least 24 times by the DCH.
A relevant background contribution to the observed \dbar signal comes from ``secondary'' \d's 
produced in nuclear interactions with the detector material
that travel inward toward the IP and are wrongly reconstructed as outward-travelling \dbar's.
To suppress this contribution we require that the transverse distance of closest approach (DOCA) of the
reconstructed trajectory to the beamspot be less than 400 \mum. 
The effect of underestimated energy loss on the measured DOCA of
tracks is found to be well reproduced in the simulation,
and we do not apply any correction to this quantity.
Finally, \d's and \dbar's in the considered momentum range are below the threshold for radiating Cherenkov light
in the DIRC quartz-glass bars, 
so we reject all the candidates with more than 10 associated Cherenkov photons
for the the best-fit DIRC mass hypothesis, $\pi$, $K$, $p$, $e$ or $\mu$.

To measure the \dbar yields, 
we apply a weight to each candidate to correct for detector and
selection acceptance, and we then extract the yields from the \dbar
candidate energy loss distributions of the DCH and SVT using a weighted fit.
Global trigger and event selection efficiencies are determined from simulated $\epem \to Y \to 2(\NN) X$ events,
where $X$ corresponds to zero or more additional final-state particles,
and $Y = \Y2S$, $\pipi\Y1S$, or $\qqbar$ in which the nucleons are produced promptly in fragmentation. These
events more closely represent the kinematics and multiplicity of signal \dbar or \d events.
Those efficiencies are assumed to be the same for \Y2S, \Y3S, and \Y4S, 
so only the first is explicitly calculated and used also for the other resonances.
Corrections for the kinematic selections are computed as a function of $p_{\rm CM}$ 
using the fraction of \dbar, in bins of CM momentum, which would pass the selection in the LAB frame.
This fraction depends on the angular distribution of \dbar's 
in the CM frame with respect to the beam axis, which we determine from MC generator
coalescence studies with coalescence momentum $p_0 = 160\mevc$ following a similar approach to~\cite{Cui:2010ud}. 
Decays of $\Upsilon \to ggg$ produce \d's and \dbar's isotropically, 
while in \continuum there is a 
dependence on the CM polar angle. 

The \dbar reconstruction efficiency is determined 
in bins of $p_{\rm LAB}$ and $\cos{\theta_{\rm LAB}}$ from the ``signal'' MC samples. 
We compute an additional correction to this efficiency to account for the differing interactions 
of \dbar and \d in the material of the \babar\ detector. 
Starting from the differing reconstruction probabilities of protons and antiprotons, as determined by {\sc Geant4}
simulation of \YnS decays, we estimate the effect of material 
interaction on \d and \dbar by rescaling for the larger 
\dbar absorption cross sections, determined in Ref.~\cite{Denisov1971253}. The survival probability for $\bar{p}$ 
follows $P_{\bar{p}} \sim e^{-\sigma_{\bar{p}} n t}$, where $n$ is the material number density and $t$ is the thickness. The corresponding probability for antideuterons is 
$P_{\bar{d}} \sim e^{-\sigma_{\bar{d}} n t} = P_{\bar{p}}^{\sigma_{\bar{d}}/\sigma_{\bar{p}}}$, therefore the required rescaling is simply related to the cross-section ratio, assumed to be constant across this momentum range.  
As a cross-check on the result, the values obtained at $\cos{\theta_{\rm LAB}} = 0$ are found to be consistent with 
a prediction obtained from \dbar inelastic cross sections from 
Ref.~\cite{PDG} and the known distribution of material in the \babar\ detector.
The final weight applied to each track is the inverse of the product of
the trigger and event selection efficiencies, the kinematic acceptance fraction, and the \dbar reconstruction efficiency.

\begin{figure}[htb]
\includegraphics[trim=0cm 0cm 0.5cm 0.8cm,clip,width=0.97\columnwidth]{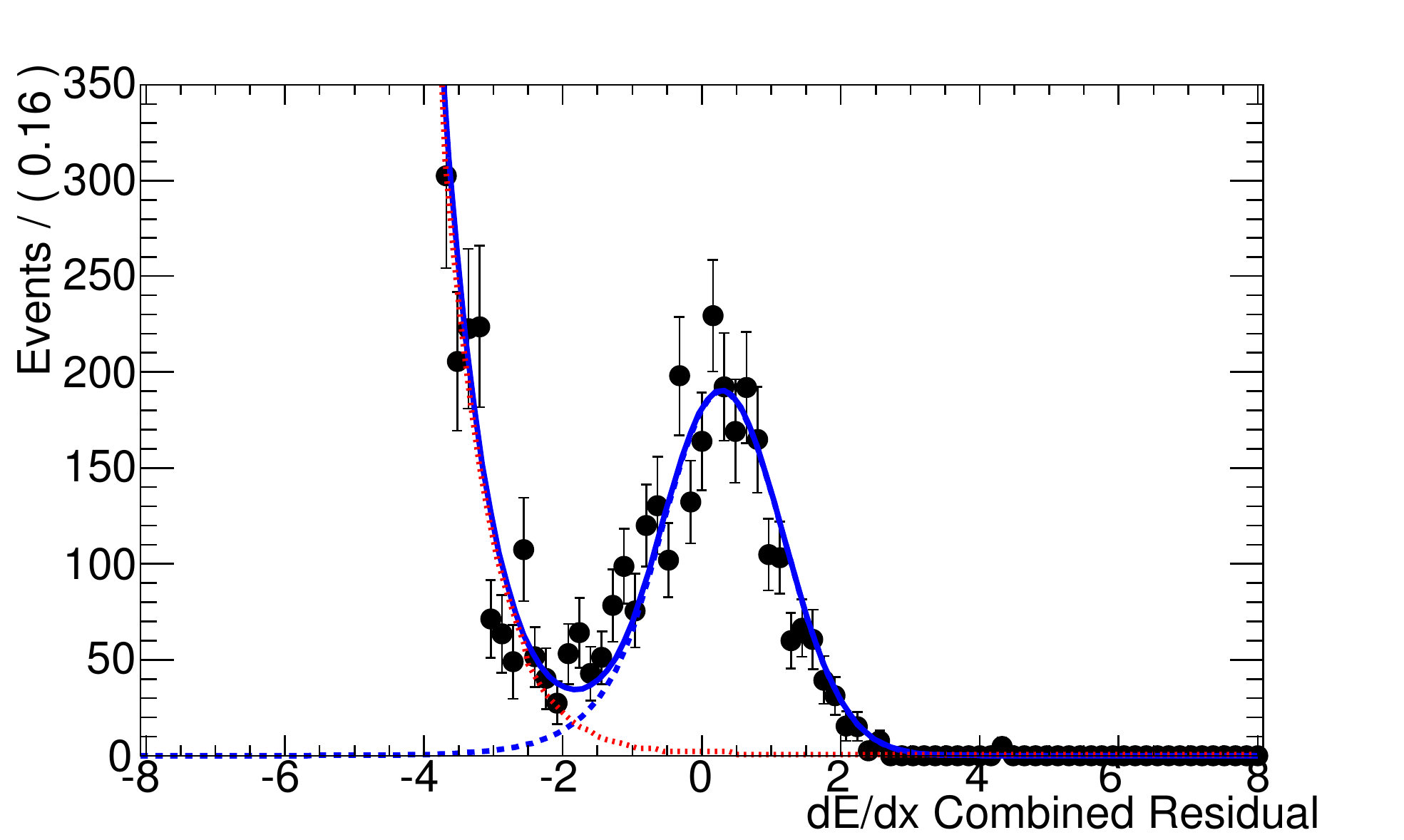}
\caption{Normalized residuals of the combined $dE/dx$ for antideuteron candidates 
in the Onpeak $\Y2S$ data sample, with fit p.d.f.'s superimposed. 
Entries have been weighted, as detailed in the text.
The solid (blue) line is the total fit, the dashed (blue) line is the
\dbar signal peak, and the dotted (red) line is the background.}
\label{fig:examplefit}
\end{figure}

Yields are computed using a fit to the distribution of the normalized residual of ionization energy loss.
Specific ionization measurements from the DCH and SVT, 
their uncertainties, and their expected values from the Bethe-Bloch formula
are calibrated using high-statistics control samples of particles of other species. 
The independent $dE/dx$ measurements, in arbitrary units, from the DCH and SVT are averaged 
according to their respective uncertainties, after rescaling the SVT measurement
and its uncertainty such that the expected value matches that of the DCH.
The normalized residual is computed as the difference between the averaged 
$dE/dx$ measurement and the expected value, 
divided by the uncertainty of the former.
Ideally, this residual (shown in Fig. \ref{fig:examplefit} for $\Y2S$ data) 
has a Gaussian distribution centered at zero for \d's and \dbar's, 
while the value for other species is far from zero (the value is positive or 
negative for particles with higher or lower mass, respectively).

The probability density function (p.d.f.) for \d's and \dbar's is estimated 
using signal MC events, while the distribution for all other particles
is taken from the distribution for negatively-charged particles for
the simulated generic decays of \YnS, which contain no true \dbar.
The signal distribution is found to be well-modeled by a piecewise combination 
of a Gaussian function with an exponential tail toward lower values. After requiring that the piecewise function and its first derivative be continuous,
the addition of the tail adds only a single parameter as compared to a pure Gaussian.
The residual distribution for other particles (``background'') is more 
complicated, and extends into the signal region. 
This background distribution, obtained from simulated $\Y2S$ events,
is described well by the sum of a Gaussian function 
and an exponential function. 
Only the functional forms of the shapes used are extracted and validated using the MC samples, 
while almost all the p.d.f.~parameters are estimated in the fit to the data. For example, the signal mean and width are constrained to be the same for \d and \dbar, hence the high-statistics sample of secondary \d determines these parameters rather than simulation.
Very few candidates with large weights are removed since they could have an undue influence on the weighted fit~\cite{James:2ndEd}.
An example of the fit to the residuals distribution for the \dbar in the \TwoS data is shown in Fig.~\ref{fig:examplefit}.
For positively charged candidates we additionally find a contribution from tritons produced in material interactions.
The tritons' distribution is similar to that of the \d's and \dbar's, 
so it is modeled using the same distribution with parameters allowed to float separately.

Candidates are divided into categories according to their charge, their CM momentum, 
and the type of dataset (Onpeak or Offpeak),
and a weighted unbinned maximum likelihood fit is performed to all categories simultaneously.
The \d, \dbar, triton and background distribution functions are the same for all categories, but the yields are floated separately.
We divide the CM momentum range [0.35,2.25]\gevc into 
nine bins containing approximately equal numbers of candidates, with no bin narrower than 100 \mevc.
To improve the quality of the fit the width of the Gaussian function for the background
is also allowed to float separately for different bins of CM momentum, to account for
significant difference in the distribution for different energy ranges.
To achieve a more stable fit, if the fit results for a split parameter 
(i.e., one allowed to take different values in different sub-samples) are statistically compatible 
between two or more sub-samples, 
the parameter is forced to have the same floating value among those sub-samples.
We perform a simultaneous fit to Onpeak and Offpeak datasets for \TwoS and \ThreeS,
obtaining the number of \dbar in each CM momentum bin.
Yield values and their uncertainties in each bin are reported in Ref.~\cite{Supp:Material}. 
We determine the final number of \dbar by subtracting the yields 
for \dbar in the Offpeak dataset from the Onpeak, 
after rescaling for the luminosity and the lowest-order $1/s$ correction of the \continuum cross section. Interference between resonant and non-resonant processes is expected to be negligible due to the small off-resonance cross-section and because on-resonance \dbar production is dominated by $\Upsilon\to ggg$ rather than $\qqbar$.

The \FourS decays almost exclusively to \BB final states, 
and \dbar production in $B$ decays is kinematically disfavored,
so the production from \FourS decays is well below our sensitivity. 
Therefore, we proceed by combining the yields from Onpeak and Offpeak datasets 
to obtain the production rate for \continuum.
As a cross-check, when subtracting the rescaled yields in the Offpeak dataset from the yields 
in the Onpeak one, the number of \dbar's are compatible with zero in all bins. 

To extract the yields in \OneS decay, we exploit candidates from $\TwoS \to \OneS \pip \pim$ decays. 
We fit the Onpeak \TwoS dataset separately in two regions of recoiling invariant mass, 
$m_{\rm recoil} = \sqrt{(E_{\rm beam} - E_{\pipi})^2 - (\vec{p}_{\rm beam} - \vec{p}_{\pipi})^2}$: 
[9.453, 9.472]\gevcc (signal) and [9.432, 9.452]\gevcc plus [9.474, 9.488]\gevcc (sidebands), 
and we subtract the yields in the sidebands, rescaled by their relative ranges, from those in the signal range.
Due to lower statistics we use only five bins in CM momentum for this measurement.

Corrections due to the requirement on the number of DCH $dE/dx$ samplings
can be computed by comparing the distribution for this variable in data and signal simulation for \d's
in a narrow signal window that provides a high-statistics control sample. This sample has negligible background due to the narrow $dE/dx$ window and the relatively large number of true deuterons produced in material.
We correct the fitted yields by $-7\%$ to correct for differences between data and simulation, which is not observed to
depend on the CM momentum or polar angle. We assume no correlation with the value of the residual itself at this level of precision.

To validate the fit procedure and check for possible biases in the \dbar yields, 
we perform a series of fits to pseudo-datasets generated according to the fit p.d.f.'s and
we assign a systematic uncertainty based on any bias found 
(i.e., the deviation from zero of the mean of the normalized residuals distribution).
The uncertainty from our choice of the background model distribution function is estimated by comparing to results obtained using
a background model consisting of two Gaussian functions fixed to a common mean.  
Additional systematic uncertainties result from the weights used to correct for
reconstruction efficiency in the detector and kinematic acceptance. To estimate the systematic uncertainty
from the event selection and trigger, we compare the nominal efficiency to that computed using a different selection where we consider only a subset of $\epem \to Y \to 2(\NN) X$ events with a $p$ or $\antiproton$ inside the nominal detector acceptance.
We evaluate the effect of finite Monte Carlo statistics by allowing the weights to
vary according to a Gaussian distribution centered at the nominal value with the width given
by the uncertainties in the weights. 
We assign uncertainties in each bin from the distribution of fit results. Any data/simulation
in tracking efficiency is known to be below the per-mille level \cite{TheBABAR:2013jta} and is negligible.
The uncertainty in the correction for \dbar material interaction is computed similarly, 
with the width of the Gaussian distribution set to the statistical 
uncertainty in the calculation plus a 30\% uncertainty in the prediction itself.

\begin{figure}[htb]
\subfigure{\begin{overpic}[trim=0cm 1.4cm 0.cm 0cm, clip, width=1.0\columnwidth]{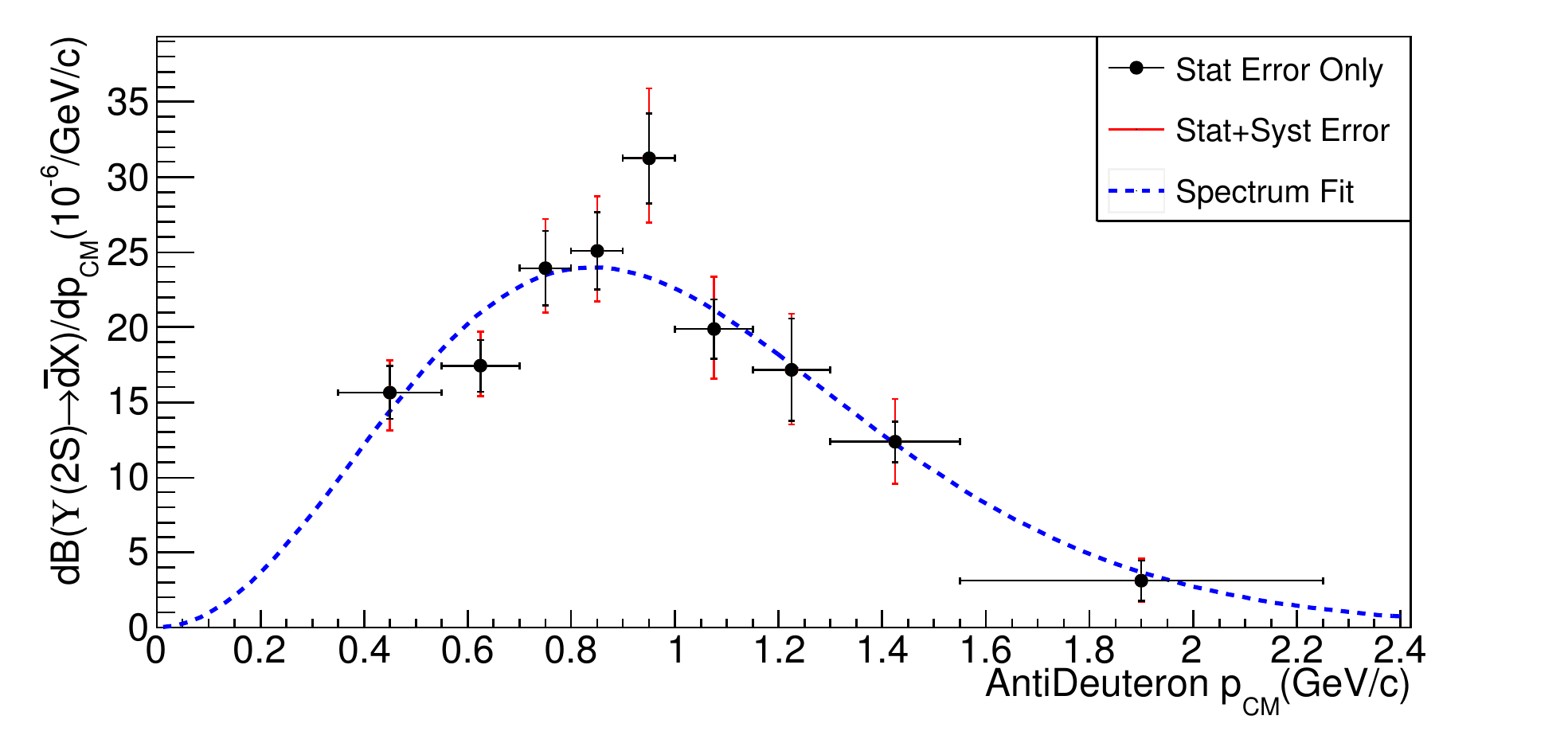}\put(62,20){(a) $\Y2S$}\end{overpic}}\vspace{-2.mm}
\subfigure{\begin{overpic}[trim=0cm 1.4cm 0.cm 0.42cm, clip, width=1.0\columnwidth]{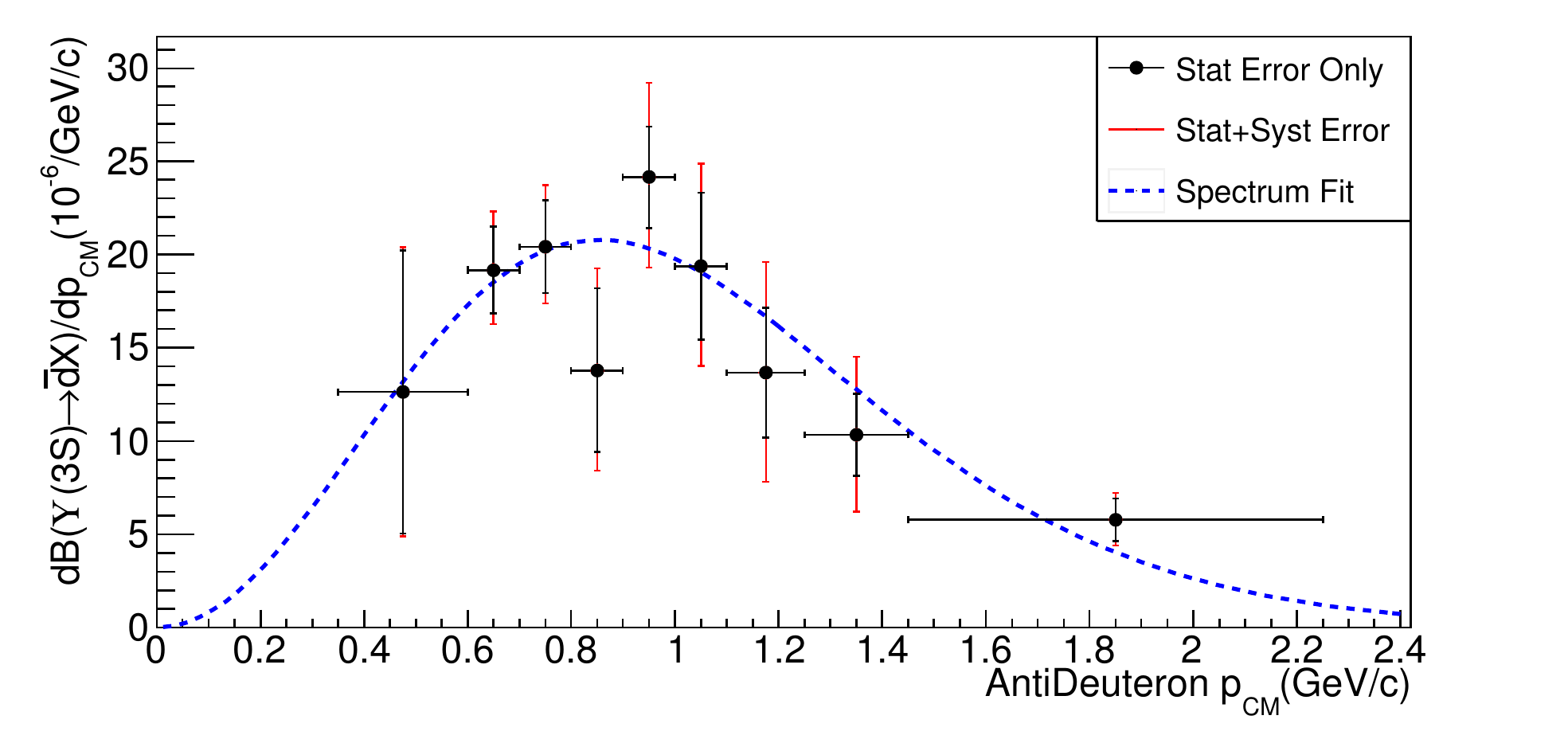}\put(62,20){(b) $\Y3S$}\end{overpic}}\vspace{-2.mm}
\subfigure{\begin{overpic}[trim=0cm 1.4cm 0.cm 0.42cm, clip, width=1.0\columnwidth]{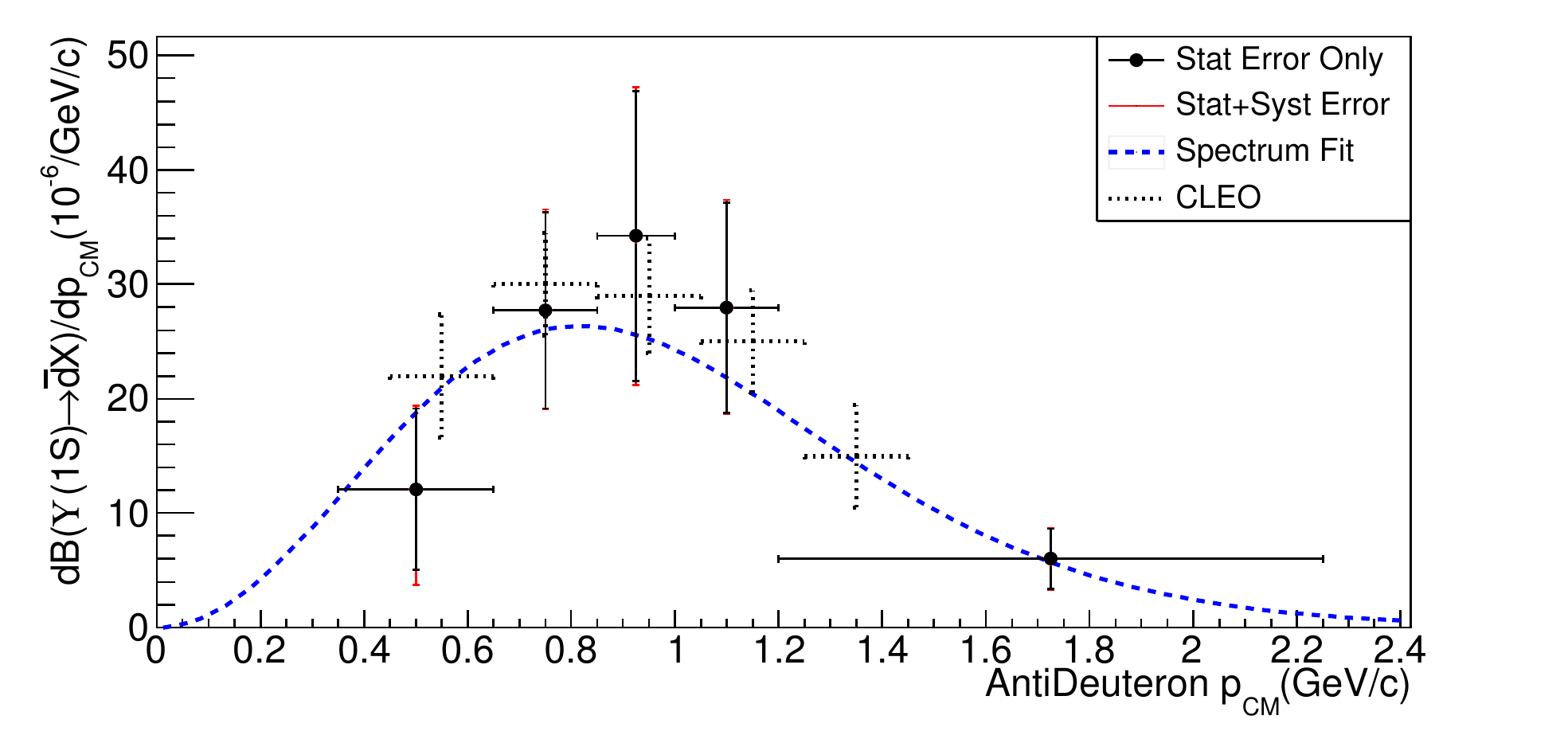}\put(62,20){(c) $\Y1S$}\end{overpic}}\vspace{-2.mm}
\subfigure{\begin{overpic}[trim=0cm 0.cm 0.cm 0.42cm, clip, width=1.0\columnwidth]{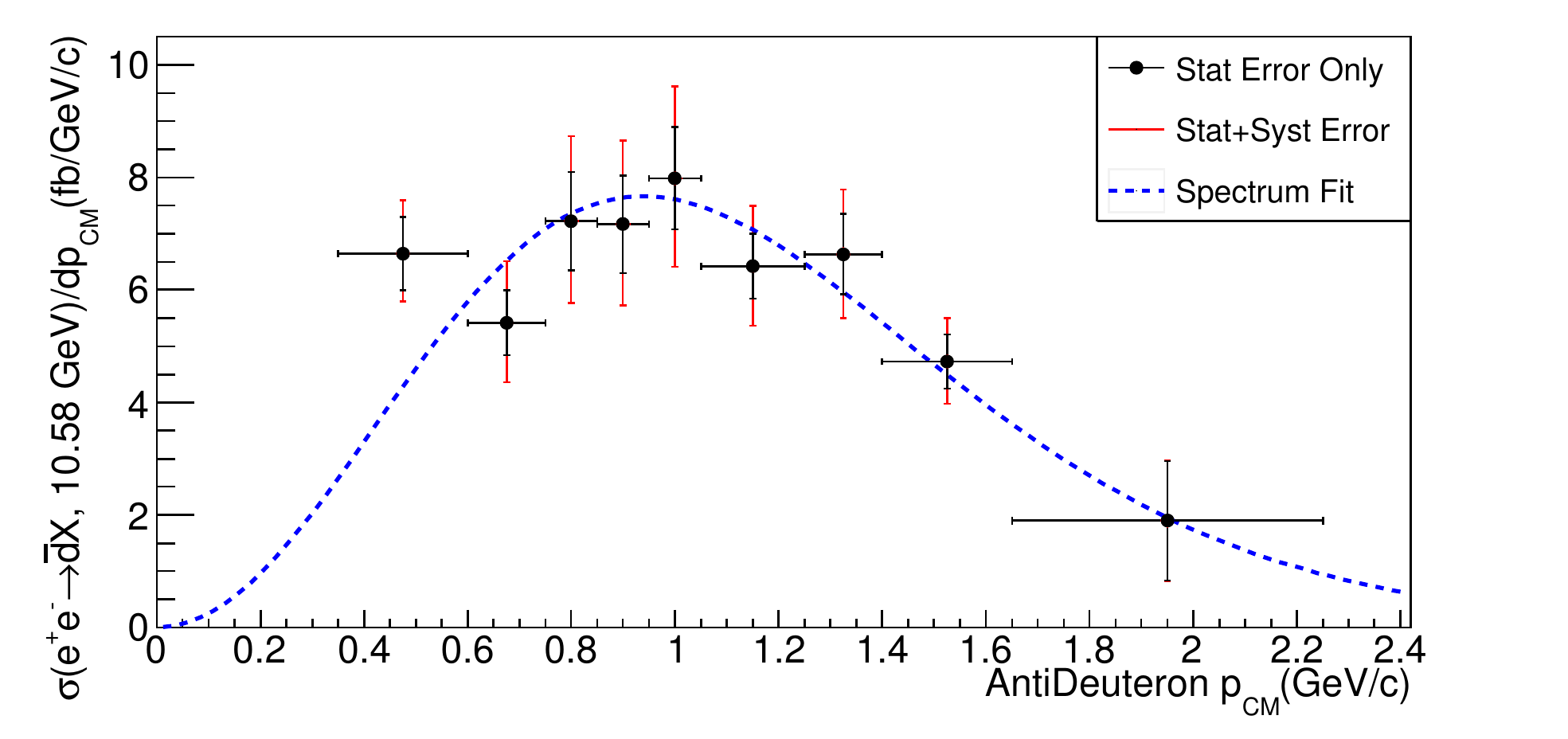}\put(62,27){(d) cont. $\epem$}\end{overpic}}
\caption{Measured antideuteron differential spectra in (a) \Y2S, (b) \Y3S, (c) \Y1S 
decays, and (d) \continuum at a CM energy of $\approx10.58$\gev.
The points with inner (black) error bars give the measurements and their associated statistical uncertainties, 
the outer (red) error bars give the quadratic sum of the statistical and systematic uncertainties, 
and the dashed (blue) curves show the fit to Eq. \ref{eq:fireball}. 
Subfig. (c) also includes the CLEO results with dashed error bars for comparison.}
\label{fig:results}
\vspace{-0.2cm}
\end{figure}

We estimate the contribution from secondary \d's, described
above, by fitting the DOCA distributions for data and simulated events
in and outside the selected region.
The simulation describes the data except for a slowly falling exponential component, 
which we ascribe to secondary \d's.  
The fits indicate a small contribution to the selected events, which we take as a one-sided systematic uncertainty.
We compare DOCA distributions of simulated antiprotons with a control sample of well-identified
antiprotons and we assign a $+5.8\%$ one-sided uncertainty.
The uncertainty in the event selection efficiency is estimated by
comparing the nominal efficiency with the efficiency for events for which at least one of the
generated prompt (anti)nucleons is a (anti)proton within the detector angular acceptance. 
Finally, uncertainties in the cross sections and number of \YnS mesons are propagated along with the other systematics. 
The contributions to the systematic uncertainties are summed in quadrature separately for the
positive and negative sides, and their values for each contribution 
and the totals are summarized in Table \ref{tab:syst}.
\begin{table*}[htb]
\caption{Contributions to the systematic uncertainties for the different measurements. 
Ranges are indicated where the contribution is different for each CM momentum bin. 
}
\begin{ruledtabular}
\begin{tabular}{ l | c c c c }
{\bf Source} & \Y2S & \Y3S & \Y1S & Continuum \\\hline
Fit Biases & 0.5\% --- 2.0\% & 0.1\% --- 6.6\% & 0.1\% --- 2.0\% & 0.0\% --- 0.2\% \\
Background Model & 0.2\% --- 7.8\% & { }{ }\!3.1\% --- 12.0\% & 0.9\% --- 7.6\% & 0.0\% --- 8.9\%\\
Reconstruction Efficiency & { }{ }\!2.5\% --- 10.5\% & { }{ }\!5.2\% --- 17.0\% & 1.3\% --- 7.1\% & 3.0\% --- 7.3\%\\
Kinematic Acceptance & { }{ }\!0.5\% --- 10.3\% & { }{ }\!3.6\% --- 16.0\% & 0.6\% --- 2.9\% & 1.4\% --- 8.4\%\\
Material Interaction & { }{ }\!2.8\% --- 10.5\% & { }{ }\!4.3\% --- 17.0\% & 2.0\% --- 7.3\% & 2.9\% --- 7.4\%\\
Fake antideuterons & $^{+0.0\%}_{-0.5\%}$ --- $^{+0.0\%}_{-9.8\%}$  & $^{+0.0\%}_{-1.1\%}$ --- $^{+0.0\%}_{-3.0\%}$ 
 & { }{ }\!$^{+0.0\%}_{-1.9\%}$ --- $^{+0.0\%}_{-32.0\%}$  & $^{+0.0\%}_{-0.6\%}$ --- $^{+0.0\%}_{-5.4\%}$ \\
DOCA Selection & $^{+5.8\%}_{-0.0\%}$  & $^{+5.8\%}_{-0.0\%}$  & $^{+5.8\%}_{-0.0\%}$  & $^{+5.8\%}_{-0.0\%}$ \\
Event Selection & 2.3\% & 2.3\% & 1.1\% & 4.6\% \\
Normalization & 1.2\% & 1.2\% & 0.2\% & 0.6\%
\end{tabular}
\end{ruledtabular}
\label{tab:syst}
\end{table*}
Systematic values in each bin are reported in Ref.~\cite{Supp:Material}.

The numbers of \dbar's extracted from the fit are corrected for
the differences between data and MC samples mentioned above, and then
are converted into branching fractions for \TwoS and \ThreeS using 
the total number of $\Upsilon$ decays.
Using the total luminosity from Onpeak and Offpeak \FourS datasets, 
we compute the observed cross section for \dbar production from \continuum at a CM energy of $\approx10.58$\gev.

The number of \OneS decays is computed as 
\begin{equation} N_{\Y1S} =  N_{\Y1S}^{\rm fit} \times \frac {f_{\rm sig} - f_{\rm sb}} {\varepsilon_{\rm filter}},\end{equation}
where $N_{\Y1S}^{\rm fit}$ is the number of $\Y2S \to \pi\pi\Y1S$ events reconstructed inclusively, 
obtained by a fit to the invariant mass 
recoiling against a reconstructed $\pipi$ system in \Y2S data. 
The quantities $f_{\rm sig}$ and $f_{\rm sb}$ are respectively the fraction of the fitted 
\Y1S recoil mass distribution in the signal and sideband regions used to subtract the contribution from background events,
and $\varepsilon_{\rm filter}$ is the average efficiency of the trigger and the event filter 
to accept $\Y2S \to \pi\pi\Y1S$ decays obtained from the \TwoS MC sample.
After applying these factors, we find $N_{\Y1S} = (9.670\pm0.023)\times10^6$.
The final values for the differential branching fractions and \continuum cross section are shown in Fig.~\ref{fig:results}.

The total rates, presented in Table \ref{tab:totalRates}, are obtained from the measured differential spectra by fits to
the ``fireball" model distribution \cite{hagedorn}
\begin{equation} \label{eq:fireball} P(E) = \alpha v^2 e^{-\beta E}
, \end{equation}
where $E$ is the \dbar CM energy
and $\alpha$ and $\beta$ are free parameters
determined by the fit. 
\begin{table}[htb]
\caption{Total rates of antideuteron production. The first uncertainties listed are statistical, the second systematic. For comparison, we also list the ratio of our measurement of the inclusive antideuteron cross section to the cross section for hadronic production at a similar energy evaluated from \cite{cleoxsec}. Here we only quote our own uncertainties, the hadronic cross section itself has a 7\% uncertainty.}
\begin{ruledtabular}
\begin{tabular}{ p{0.56\columnwidth}  p{0.40\columnwidth} }
{\bf Process} &{\bf Rate} \\\hline
$\mathcal{B}(\Y3S \to \bar{d}X)$ & \threeSresultSm \\
$\mathcal{B}(\Y2S \to \bar{d}X)$ & \twoSresultSm \\
$\mathcal{B}(\Y1S \to \bar{d}X)$ & \oneSresultSm \\
$\sigma (\epem \to \bar{d}X)$ $[\sqrt{s}\approx 10.58\gev]$ & \contresultSm \\
[1.5ex] $\displaystyle \frac{\sigma(\epem \to \bar{d}X)}{\sigma(\epem \to {\rm Hadrons})}$ & $(3.01 \pm 0.13{}^{+0.37}_{-0.31})\! \times \! 10^{-6}$\\
\end{tabular}
\end{ruledtabular}
\label{tab:totalRates}
\end{table}
The fits are shown in Fig.~\ref{fig:results}.
The total rates quoted in Table \ref{tab:totalRates} are the integral of these distributions from 0\gevc to 4\gevc, and the
associated uncertainties are those in the integral taking into account the full covariance
of $\alpha$ and $\beta$. We find that values for the parameter $\beta$ in $\Upsilon$ decays 
are mutually compatible within $1\sigma$, and the average value is $\beta = (4.71\pm0.19)\gev^{-1}$. 
Fitting to the \continuum spectrum yields a lower value of $(3.92\pm0.22)\gev^{-1}$, 
corresponding to a somewhat harder spectrum.
As an additional cross-check on the cross section for \continuum production, 
we fit separately the differential spectra from the \Y4S Offpeak dataset only,
and obtain a cross section of $12.2 \pm 1.8 \fb$, where the error is statistical only.
Values for both this cross section and other parameters of the fit
are in good agreement with the Onpeak plus Offpeak result.

In summary, we have performed measurements of inclusive \dbar production
in $\Upsilon(1,2,3S)$ decays and in \continuum. 
These are the first measurements of \dbar production in 
\continuum at a CM energy of $\approx10.58$\gev and in $\Y3S$ decay, and the most precise measurement in $\Y2S$ decay. 
Our total and differential rates for inclusive $\bar{d}$
production in \Y1S and \Y2S decay are in good agreement with previous 
measurements \cite{PDG}, and with the expected spectral shapes from the coalescence model~\cite{coamodel,coamodel1}. 
We additionally note an order of magnitude suppression of \dbar production 
in quark-dominated \continuum relative to the gluon-dominated $\Upsilon$ decays.

We are grateful for the excellent luminosity and machine conditions
provided by our \pep2 colleagues, 
and for the substantial dedicated effort from
the computing organizations that support \babar.
The collaborating institutions wish to thank 
SLAC for its support and kind hospitality. 
This work is supported by
DOE
and NSF (USA),
NSERC (Canada),
CEA and
CNRS-IN2P3
(France),
BMBF and DFG
(Germany),
INFN (Italy),
FOM (The Netherlands),
NFR (Norway),
MES (Russia),
MICIIN (Spain),
STFC (United Kingdom). 
Individuals have received support from the
Marie Curie EIF (European Union),
the A.~P.~Sloan Foundation (USA)
and the Binational Science Foundation (USA-Israel).

\end{document}


\title{
{
\large \boldmath
Supplemental Material for \\
Antideuteron production  in \YnS  decays and in \continuum}
}

\begin{abstract}
Supplemental material for paper: xxxx.
\end{abstract}

\maketitle

\renewcommand*{\arraystretch}{1.5}

\begin{table*}[htb]
\caption{\tabcap{\Y2S decays}}
\label{tab:dbarsyst2s}
\centering
\begin{tabular}{l|r|r|r|r|r|r|r|r|r}\hline \hline
Monmentum range (\gevc)                          & 0.35 -- 0.55 & 0.70 & 0.80 & 0.90 & 1.00 & 1.15 & 1.30 & 1.55 & 2.25 \\
\hline
Events in Signal Region                  & 81.0 & 105.0 & 97.0 & 93.0 & 103.0 & 95.0 & 67.8 & 71.4 & 57.4 \\
\hline
Branching Ratio ($\times 10^{-6} / (\gevc)$) & 15.66 & 17.43 & 23.93 & 25.07 & 31.25 & 19.89 & 17.16 & 12.35 & 3.13 \\
\hline
Statistical error (\%)                           & 10.49 & 9.21 & 9.62 & 9.58 & 8.92 & 9.29 & 18.51 & 10.30 & 40.16 \\
\hline
Fit Biases                & 0.62 & 0.65 & 1.11 & 1.11 & 0.50 & 0.51 & 2.27 & 0.55 & 3.99 \\
Background Model          & 0.30 & 0.32 & 0.16 & 0.52 & 0.65 & 0.90 & 0.27 & 0.23 & 7.72 \\
Reconstruction Efficiency & 2.64 & 2.52 & 3.02 & 4.10 & 4.86 & 6.67 & 2.84 & 10.47 & 6.47 \\
Kinematic Acceptance      & 0.53 & 2.16 & 2.51 & 3.91 & 4.68 & 6.76 & 2.74 & 10.33 & 5.92 \\
Material Interaction      & 2.79 & 2.98 & 3.26 & 4.30 & 5.04 & 7.01 & 2.79 & 10.48 & 6.65 \\
Fake antideuterons        & -9.77 & -2.09 & -2.14 & -1.66 & -2.31 & -3.30 & -3.43 & -1.73 & -0.45 \\
DOCA Selection            & +5.82 & +5.82 & +5.82 & +5.82 & +5.82 & +5.82 & +5.82 & +5.82 & +5.82 \\
Event Selection           & 2.30 & 2.30 & 2.30 & 2.30 & 2.30 & 2.30 & 2.30 & 2.30 & 2.30 \\
Normalization             & 1.20 & 1.20 & 1.20 & 1.20 & 1.20 & 1.20 & 1.20 & 1.20 & 1.20 \\
\hline
Total systematic error (\%) & $^{ +7.49}_{ -10.85}$ & $^{ +7.82}_{ -5.62}$ & $^{ +8.24}_{ -6.21}$ & $^{ +9.63}_{ -7.84}$ & $^{ +10.59}_{ -9.15}$ & $^{ +13.46}_{ -12.57}$ & $^{ +8.32}_{ -6.86}$ & $^{ +19.16}_{ -18.34}$ & $^{ +15.40}_{ -14.27}$ \\
\hline
Total error (\%) & $^{ +12.89}_{ -15.09}$ & $^{ +12.08}_{ -10.79}$ & $^{ +12.67}_{ -11.45}$ & $^{ +13.58}_{ -12.38}$ & $^{ +13.85}_{ -12.78}$ & $^{ +16.35}_{ -15.63}$ & $^{ +20.29}_{ -19.74}$ & $^{ +21.76}_{ -21.03}$ & $^{ +43.01}_{ -42.62}$ \\
\hline \hline
\end{tabular}
\end{table*}

\begin{table*}[htb]
\caption{\tabcap{\Y3S decays}}
\label{tab:3ssyst}
\centering
\begin{tabular}{l|r|r|r|r|r|r|r|r|r}\hline \hline
Bin range (\gevc)  & 0.35 -- 0.60 & 0.70 & 0.80 & 0.90 & 1.00 & 1.10 & 1.25 & 1.45 & 2.25 \\
\hline
Events in Signal Region & 89.2 & 93.0 & 92.0 & 61.5 & 91.0 & 70.2 & 75.5 & 71.2 & 118.5 \\
\hline
Branching Ratio ($\times 10^{-6} / (\gevc)$) & 12.63 & 19.18 & 20.42 & 13.80 & 24.14 & 19.39 & 13.68 & 10.34 & 5.78 \\
\hline
Statistical error (\%) & 56.04 & 11.25 & 11.27 & 29.51 & 10.48 & 18.83 & 23.69 & 19.63 & 18.46 \\
\hline
Fit Biases                & 2.78 & 0.10 & 0.70 & 9.57 & 0.64 & 3.41 & 6.55 & 5.31 & 4.89 \\
Background Model          & 3.09 & 3.40 & 3.57 & 5.61 & 7.44 & 5.35 & 12.45 & 11.74 & 6.74 \\
Reconstruction Efficiency & 5.19 & 4.22 & 3.41 & 11.66 & 7.78 & 9.34 & 16.93 & 16.99 & 7.07 \\
Kinematic Acceptance      & 4.42 & 3.62 & 3.59 & 10.82 & 7.34 & 9.38 & 16.34 & 16.39 & 4.85 \\
Material Interaction      & 5.64 & 4.71 & 4.27 & 11.43 & 7.64 & 9.45 & 16.78 & 16.45 & 4.92 \\
Fake antideuterons        & -2.90 & -1.51 & -1.07 & -1.62 & -1.99 & -1.59 & -2.94 & -2.04 & -0.97 \\
DOCA Selection            & +5.82 & +5.82 & +5.82 & +5.82 & +5.82 & +5.82 & +5.82 & +5.82 & +5.82 \\
Event Selection           & 2.30 & 2.30 & 2.30 & 2.30 & 2.30 & 2.30 & 2.30 & 2.30 & 2.30 \\
Normalization             & 1.20 & 1.20 & 1.20 & 1.20 & 1.20 & 1.20 & 1.20 & 1.20 & 1.20 \\
\hline
Total systematic error (\%) & $^{ +11.66}_{ -10.52}$ & $^{ +10.26}_{ -8.58}$ & $^{ +9.83}_{ -7.99}$ & $^{ +23.40}_{ -22.72}$ & $^{ +16.41}_{ -15.47}$ & $^{ +18.58}_{ -17.72}$ & $^{ +32.77}_{ -32.38}$ & $^{ +32.17}_{ -31.70}$ & $^{ +14.41}_{ -13.22}$ \\
\hline
Total error (\%) & $^{ +57.24}_{ -57.02}$ & $^{ +15.23}_{ -14.15}$ & $^{ +14.95}_{ -13.81}$ & $^{ +37.66}_{ -37.24}$ & $^{ +19.47}_{ -18.69}$ & $^{ +26.46}_{ -25.86}$ & $^{ +40.44}_{ -40.12}$ & $^{ +37.68}_{ -37.29}$ & $^{ +23.42}_{ -22.71}$ \\
\hline\hline
\end{tabular}
\end{table*}

\begin{table*}[htb]
\caption{\tabcap{\Y1S decays}}
\label{tab:dbarsyst1s}
\centering
\begin{tabular}{l|r|r|r|r|r}\hline \hline
Bin range (\gevc)  & 0.35 -- 0.65 & 0.85 & 1.00 & 1.20 & 2.25 \\
\hline
Events in Signal Region & 11.6 & 22.6 & 18.9 & 18.3 & 19.5 \\
\hline
Branching Ratio ($\times 10^{-6} / (\gevc)$) & 12.09 & 27.73 & 34.22 & 27.98 & 6.00 \\

Statistical error (\%) & 54.14 & 28.75 & 34.42 & 30.55 & 40.43 \\
\hline
Fit Biases                & 1.16 & 0.07 & 0.22 & 0.31 & 2.03 \\
Background Model          & 7.58 & 0.86 & 2.22 & 1.65 & 3.83 \\
Reconstruction Efficiency & 7.08 & 1.68 & 3.57 & 1.28 & 3.23 \\
Kinematic Acceptance      & 2.36 & 0.59 & 2.49 & 0.75 & 2.90 \\
Material Interaction      & 7.27 & 2.47 & 2.84 & 1.95 & 2.97 \\
Fake antideuterons        & -32.00 & -1.90 & -5.52 & -3.46 & -4.85 \\
DOCA Selection            & +5.82 & +5.82 & +5.82 & +5.82 & +5.82 \\
Event Selection           & 1.11 & 1.11 & 1.11 & 1.11 & 1.11 \\
Normalization             & 0.24 & 0.24 & 0.24 & 0.24 & 0.24 \\
\hline
Total systematic error (\%) & $^{ +14.22}_{ -34.53}$ & $^{ +6.72}_{ -3.86}$ & $^{ +8.20}_{ -7.99}$ & $^{ +6.63}_{ -4.70}$ & $^{ +9.04}_{ -8.44}$ \\
\hline
Total error (\%) & $^{ +55.97}_{ -64.21}$ & $^{ +29.52}_{ -29.01}$ & $^{ +35.39}_{ -35.34}$ & $^{ +31.26}_{ -30.91}$ & $^{ +41.42}_{ -41.30}$ \\
\hline\hline
\end{tabular}
\end{table*}

\begin{table*}[htb]
\caption{\tabcap{\Y4S}}
\label{tab:4ssyst}
\centering
\begin{tabular}{l|r|r|r|r|r|r|r|r|r}\hline \hline
Bin range (\gevc)  & 0.35 -- 0.60 & 0.75 & 0.85 & 0.95 & 1.05 & 1.25 & 1.40 & 1.65 & 2.25 \\
\hline
Events in Signal Region & -25.5 & 1.20 & -10.1 & -16.1 & -22.5 & -27.9 & -11.5 & 27.7 & 39.7 \\
\hline
Branching Ratio ($\times 10^{-6} / (\gevc)$) & 1.17 & -1.06 & -2.42 & -3.62 & -3.79 & -1.69 & -2.05 & 0.39 & 1.72 \\
\hline
Statistical error (\%) & 161.00 & 222.17 & 142.97 & 100.48 & 105.69 & 149.04 & 150.48 & 445.82 & 556.71 \\
\hline
Fit Biases                & 4.64 & 16.06 & 10.40 & 4.72 & 4.53 & 6.56 & 2.38 & 32.11 & 0.35 \\
Background Model          & 4.06 & 14.91 & 7.49 & 3.85 & 0.70 & 6.81 & 8.98 & 2.12 & 1.56 \\
Reconstruction Efficiency & 71.78 & 122.04 & 86.46 & 61.57 & 57.38 & 62.42 & 62.28 & 187.82 & 3.35 \\
Kinematic Acceptance      & 68.12 & 120.62 & 84.95 & 62.45 & 56.45 & 61.11 & 60.97 & 187.15 & 3.25 \\
Material Interaction      & 74.08 & 168.33 & 89.82 & 64.72 & 59.46 & 63.95 & 64.36 & 198.58 & 3.51 \\
Fake antideuterons        & -1.02 & -3.91 & -1.96 & -1.90 & -1.99 & -6.13 & -5.60 & -16.17 & -0.61 \\
DOCA Selection            & +5.82 & +5.82 & +5.82 & +5.82 & +5.82 & +5.82 & +5.82 & +5.82 & +5.82 \\
Event Selection           & 2.30 & 2.30 & 2.30 & 2.30 & 2.30 & 2.30 & 2.30 & 2.30 & 2.30 \\
Normalization             & 0.60 & 0.60 & 0.60 & 0.60 & 0.60 & 0.60 & 0.60 & 0.60 & 0.60 \\
\hline
Total systematic error (\%) & $^{ +123.93}_{ -123.79}$ & $^{ +241.45}_{ -241.41}$ & $^{ +151.54}_{ -151.44}$ & $^{ +109.34}_{ -109.21}$ & $^{ +100.37}_{ -100.23}$ & $^{ +108.85}_{ -108.87}$ & $^{ +108.92}_{ -108.91}$ & $^{ +332.88}_{ -333.22}$ & $^{ +8.73}_{ -6.53}$ \\
\hline
Total error (\%) & $^{ +203.17}_{ -203.09}$ & $^{ +328.11}_{ -328.08}$ & $^{ +208.34}_{ -208.27}$ & $^{ +148.50}_{ -148.40}$ & $^{ +145.76}_{ -145.66}$ & $^{ +184.56}_{ -184.57}$ & $^{ +185.77}_{ -185.76}$ & $^{ +556.39}_{ -556.59}$ & $^{ +556.78}_{ -556.75}$ \\
\hline\hline
\end{tabular}
\end{table*}

\begin{table*}[htb]
\caption{\tabcap{continuum \epem annihilation}}
\label{tab:dbarsystcont}
\centering
\begin{tabular}{l|r|r|r|r|r|r|r|r|r}\hline \hline
Bin range (\gevc)  & 0.35 -- 0.60 & 0.75 & 0.85 & 0.95 & 1.05 & 1.25 & 1.40 & 1.65 & 2.25 \\
\hline
Events in Signal Region & 157 & 130 & 108 & 102 & 117 & 176 & 128 & 178 & 308 \\
\hline
Cross Section ($0.1 \rm{fb}/(\gevc)$) & 66.46 & 54.13 & 72.22 & 71.70 & 79.84 & 64.24 & 66.38 & 47.28 & 18.97 \\
\hline
Statistical error (\%) & 9.05 & 9.97 & 11.16 & 11.28 & 10.63 & 8.34 & 10.05 & 9.44 & 51.94 \\
\hline
Fit Biases                & 0.05 & 0.05 & 0.18 & 0.07 & 0.02 & 0.02 & 0.06 & 0.00 & 0.19 \\
Background Model          & 0.22 & 8.86 & 5.05 & 3.61 & 6.91 & 6.43 & 5.23 & 6.40 & 0.04 \\
Reconstruction Efficiency & 3.77 & 6.28 & 6.95 & 7.34 & 6.46 & 4.22 & 5.35 & 4.88 & 3.02 \\
Kinematic Acceptance      & 1.47 & 6.17 & 8.32 & 8.40 & 7.69 & 5.93 & 4.08 & 1.37 & 4.17 \\
Material Interaction      & 4.57 & 6.71 & 7.29 & 7.36 & 6.80 & 4.78 & 5.47 & 4.80 & 2.88 \\
Fake antideuterons        & -0.56 & -2.37 & -2.04 & -2.98 & -2.93 & -5.01 & -5.37 & -4.15 & -1.73 \\
DOCA Selection            & +5.82 & +5.82 & +5.82 & +5.82 & +5.82 & +5.82 & +5.82 & +5.82 & +5.82 \\
Event Selection           & 4.61 & 4.61 & 4.61 & 4.61 & 4.61 & 4.61 & 4.61 & 4.61 & 4.61 \\
Normalization             & 0.60 & 0.60 & 0.60 & 0.60 & 0.60 & 0.60 & 0.60 & 0.60 & 0.60 \\
\hline
Total systematic error (\%) & $^{ +9.64}_{ -7.70}$ & $^{ +16.02}_{ -15.11}$ & $^{ +15.86}_{ -14.90}$ & $^{ +15.72}_{ -14.91}$ & $^{ +15.83}_{ -15.00}$ & $^{ +13.15}_{ -12.81}$ & $^{ +12.57}_{ -12.37}$ & $^{ +12.05}_{ -11.34}$ & $^{ +9.50}_{ -7.71}$ \\
\hline
Total error (\%) & $^{ +13.22}_{ -11.88}$ & $^{ +18.87}_{ -18.10}$ & $^{ +19.40}_{ -18.62}$ & $^{ +19.35}_{ -18.69}$ & $^{ +19.06}_{ -18.39}$ & $^{ +15.57}_{ -15.28}$ & $^{ +16.09}_{ -15.94}$ & $^{ +15.31}_{ -14.75}$ & $^{ +52.80}_{ -52.51}$ \\
\hline\hline
\end{tabular}
\end{table*}